\documentstyle[twoside,fleqn,espcrc2,graphicx,here]{article}

\newcommand{\AmS}{{\protect\the\textfont2
  A\kern-.1667em\lower.5ex\hbox{M}\kern-.125emS}}
\newcommand {\Figref}[1]{Figure~\ref{fig:#1}}

\title{Study of Multi-muon Events from EAS with the L3 Detector
       at Shallow Depth Underground}

\author{D. Bourilkov
        \address{Institute for Particle Physics (IPP), ETH Z\"urich,
        CH-8093 Z\"urich, Switzerland}%
        \thanks{\tt \mbox{e-mail: Dimitri.Bourilkov@cern.ch}}
        (for the L3+C Collaboration)}
       
\begin{document}
\begin{titlepage}
\begin{flushright} 
{\large
astro-ph/0002326 \\
February 16, 2000
}
\end{flushright}

\vspace*{3.0cm}

\begin{center} {\Large \bf
       Study of Multi-muon Events from EAS with the L3 Detector\\
\vspace*{0.15cm}
       at Shallow Depth Underground}

\vspace*{2.0cm}
  {\Large
  Dimitri Bourilkov\footnote{\tt e-mail: Dimitri.Bourilkov@cern.ch}
  (for the L3+C Collaboration)}

\vspace*{1.0cm}
{\large
  Institute for Particle Physics (IPP), ETH Z\"urich, \\
  CH-8093 Z\"urich, Switzerland
}

\vspace*{1.0cm}
{\Large
Presented at the 6th International Workshop on\\
\vspace*{0.11cm}
Topics in Astroparticle and Underground Physics\\
\vspace*{0.24cm}
TAUP99, September 6-10, 1999, Paris, France
}
\vspace*{3.3cm}
\end{center}

%
%
%

\vspace*{1.0cm}

\end{titlepage}
\clearpage
\renewcommand{\thefootnote}{\arabic{footnote}}
\setcounter{footnote}{0}

\begin{abstract}
We present first preliminary data from the L3+Cosmics
experiment and results from Monte Carlo simulations of multi-muon events
as observed 30 m underground.
\end{abstract}

\maketitle

\section{THE L3+COSMICS EXPERIMENT}
The muon component of extensive air showers (EAS), due to the long muon range
in the Earth's atmosphere, carries a wealth of information about the
shower development. Study of multi-muon events gives an insight into the
primary cosmic ray composition and the physics of high energy hadronic
interactions.
The L3 detector, situated 30 m underground, offers
interesting possibilities to detect and study such
events~\cite{PL97}, which are
complementary to the data collected in traditional cosmic ray
experiments.
The hadron component of EAS is absorbed, while the muon component is
detected with low threshold (typically, if we exclude access shafts,
15 GeV) and high momentum and spatial resolution by the sophisticated
tracking system of the L3 detector. The muon spectrum can be measured
up to 2 TeV with high precision. The multi-muon event rate is high enough
to make studies of the knee region possible with one year of data taking.

This year, 5 billion triggers were collected with the full L3+Cosmics
setup. The independent readout and data acquisition system allows us
to take data in parallel with L3. The acceptance of the setup is
200~$\rm m^2 sr$. The angular resolution is better than 3.5 mrad for
muons above 100 GeV and zenith angles from 0 to $\rm 50^{\circ}$.
The momentum resolution is 5.0 \% at
45 GeV. It is calibrated with $\rm Z \rightarrow \mu^+\mu^-$ events from
the LEP calibration runs, where the muon momentum is known exactly.

\section{MONTE CARLO SIMULATIONS}
The simulation program ARROW~\cite{DB91,DB99} is used to calculate the
hadron, muon and neutrino flux at the detector level. The method
combines simulations for fixed energies and different primary nuclei
with a parametrization of the energy dependence and allows to do
fast calculations for different geometries and energy thresholds.
Results for the L3+C setup with sensitive area 200~$\rm m^2 sr$
\begin{figure}[htb]
\vspace{-0.3cm}
\resizebox{8.1cm}{6.8cm}{\includegraphics*[1.5cm,2.0cm][15cm,10cm]{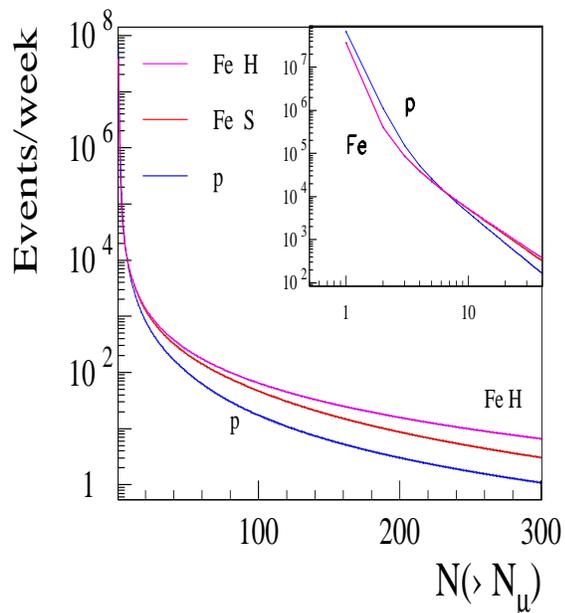}}
\caption{Integrated $\mu$ multiplicity - number of expected events with
           $\rm N(>N_{\mu})$ for a week of data taking.
           Lower curve - protons,
           middle curve - iron S,
           upper curve - iron H (see text).
           The predicted rate agrees well with the experimental one.}
\label{fig:nmu2}
\end{figure}
are shown in~\Figref{nmu2}.
\begin{figure*}[htb]
\vspace{9pt}
\resizebox{\textwidth}{11cm}{\includegraphics*[1.0cm,6.5cm][21cm,24cm]{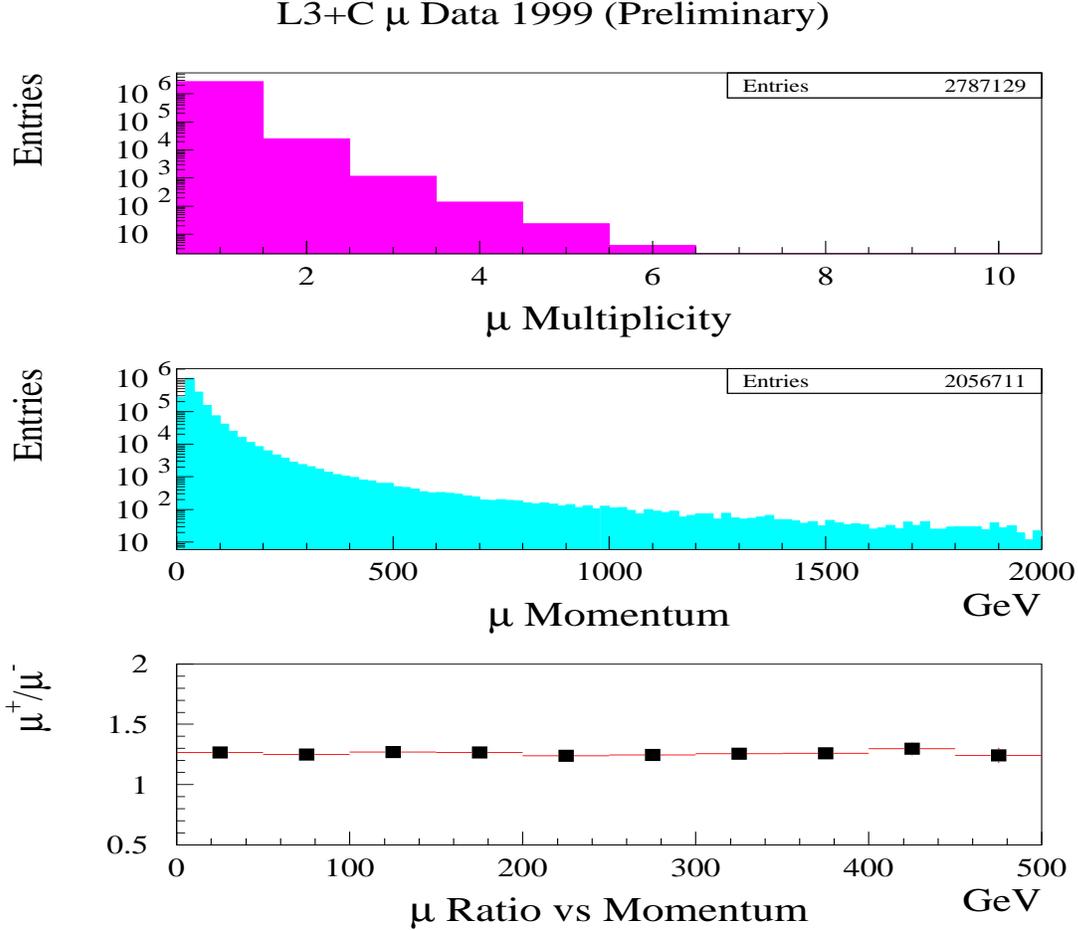}}
\caption{Preliminary data from the 1999 run. No corrections for acceptance
or efficiency are applied.}
\label{fig:l3c}
\end{figure*}
The primary composition is divided in heavy (Fe) and
light (p) components in two limiting hypotheses:
Fe S for constant heavy contribution $\rm \sim 30$\% and
Fe H for heavy contribution rising from 30\% below to 70\%
above the knee. The events with up to 6 muons are dominated by
proton induced showers and above 10 muons the iron takes over.
To distinguish between the two hypotheses  with this method
we need to detect events with $\sim 50$ muons and more.

\section{DATA AND OUTLOOK}
A first small subset of our data is shown in~\Figref{l3c}.
Only part of the events with up to 6 muons are reconstructed currently.
The observed charge ratio $\mu^+/\mu^-$ from the raw data
is flat in the region between 50 and 500 GeV.

In the year 2000 an EAS array will be mounted above L3+C in order
to detect the primary energy and core position. 
The experimental program includes studies of 
muon families and the primary composition,
sidereal anisotropies,
high multiplicity events in coincidences with other experiments,
the moon shadow,
searches for point sources,
gamma ray bursts
and exotic events.

\end{document}